# Permanent magnets based on hard ferrite ceramics

Cecilia Granados-Miralles,[a]* Matilde Saura-Múzquiz,[b] and Henrik L. Andersen[b]

[a] *Departamento de Electrocerámica, Instituto de Cerámica y Vidrio, CSIC, 28049 Madrid, Spain*
[b] *Departamento de Física de Materiales, Universidad Complutense de Madrid, 28040 Madrid, Spain*
*c.granados.miralles@icv.csic.es

## Abstract

Permanent magnets are integral components in many of the modern technologies that are critical for the transition to a sustainable society. However, most of the high-performance ($BH_{max} > 100$ kJ/m³) permanent magnets that are currently employed contain rare-earth elements (REE), which have long been classified as critical materials with a high supply risk and concerns regarding pollution in their mining. Therefore, suitable REE-lean/free magnets must be developed in order to ensure the sustainability of clean energy generation and electric mobility. The REE-free hexagonal ferrites (or hexaferrites) are the most used permanent magnets across all applications, with an 85 wt.% pie of the permanent magnet market. They are the dominant lower-grade option ($BH_{max} < 25$ kJ/m³) due to their relatively good hard magnetic properties, high Curie temperature (>700 K), low cost and good chemical stability. In recent years, the hexaferrites have also emerged as candidates for substituting REE-based permanent magnets in applications requiring intermediate magnetic performance (25–100 kJ/m³), due to considerable performance improvements achieved through chemical tuning, nanostructuring and compaction/sintering optimization. This chapter reviews the state-of-the-art sintering strategies being investigated with the aim of manufacturing hexaferrite magnets with optimized magnetic properties, identifying key challenges and highlighting the natural future steps to be followed.

# 1. Introduction

## 1.1 Classification of magnetic materials

The magnetism of magnetic materials arises at the atomic scale and is influenced by characteristics spanning several orders of magnitude (see Figure 1a). In the atoms of most compounds, the electrons exist in pairs with opposite spins that cancel out each other's magnetic moment. However, some elements or ions have unpaired electrons, whose spin and orbital motion cause them to exhibit a magnetic field giving the atom a magnetic moment. The organization of these magnetic atoms in the atomic structure of the material determines its magnetic properties. Figure 1b shows a schematic illustration of the main types of magnetic ordering. In paramagnetic materials, the atomic magnetic moments are randomly oriented leading to no net magnetization and a relatively weak attraction to an external magnetic field. Antiferromagnetic materials are magnetically ordered, but also exhibit zero net magnetization due to an antiparallel organization of equal atomic magnetic moments. However, in ferro- or ferri-magnetic materials (below their Curie temperature, $T_c$, which is the critical temperature above which thermal fluctuations lead to the material being paramagnetic), the magnetic atoms are organized in a way that leads to a net magnetization along a certain direction (magnetic easy axis) in the structure, and it is these types of materials that are used for permanent magnets (PMs).

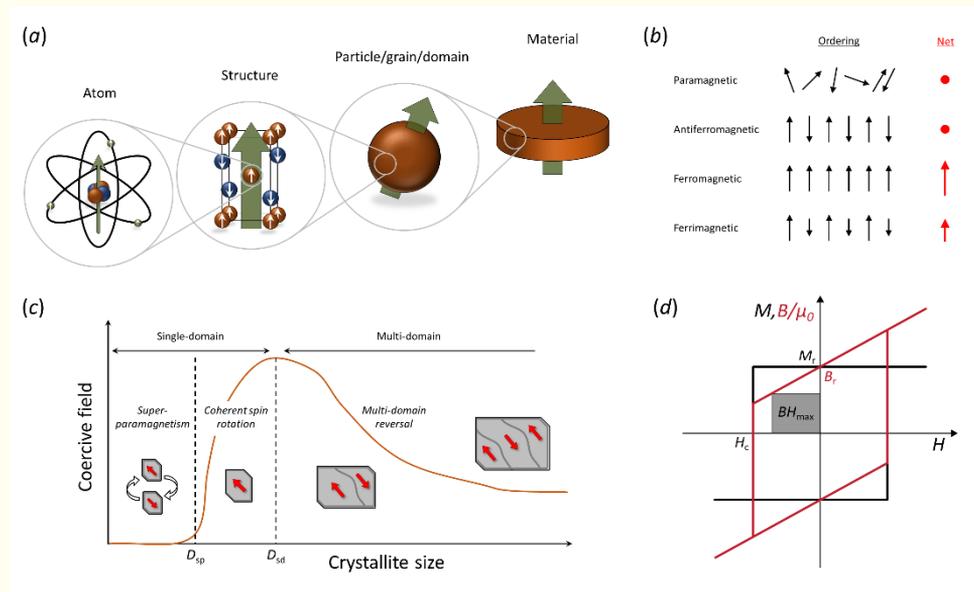

Figure 1. (a) Illustration of the multiscale origin of the magnetism in magnetic materials. (b) Schematic illustration of main magnetic ordering types and the resulting net zero field magnetization. (c) Size dependency of the coercivity. (d) Hysteresis curve of an ideal permanent magnet.

The ferro/ferri-magnetic materials are generally categorized as either 'soft' or 'hard' depending on their resistance to demagnetization. This is evaluated in terms of the coercive field (or coercivity, $H_c$), which is the external magnetic field required to reset the magnetization of the material. Magnetically soft materials are easily (de)magnetized by an external magnetic field (typically defined as $H_c < 10$ kA/m) and their magnetization is therefore often temporary, while hard (or permanent) magnetic materials have a high resistance to demagnetization ($H_c > 400$ kA/m) and once magnetized they can therefore sustain a magnetic field indefinitely.[1] The coercivity of a material is determined in part by the intrinsic magnetocrystalline anisotropy of the crystal structure as well as by microstructural (extrinsic) effects such as crystallite size or structural defects, which influence the formation (nucleation and growth) of magnetic domains in the material. For most magnetic materials, $H_c$ is found to increase as the crystallite size is reduced, reaching a maximum value at the critical single-domain size (see Figure 1c).

Another key property of a magnetic material is its remanence field ($B_r$ or $M_r$), which is the spontaneous magnetic flux density or magnetization exhibited by the material in zero external field conditions. Figure 1d shows a schematic illustration of the external magnetic field ($H$)-dependent flux density ($B$) and magnetization ($M$) curves, commonly called hysteresis curves, of an ideal permanent magnetic material. As illustrated, it is the combination of these two parameters, *i.e.*, the coercivity (magnetic stability) and remanence (spontaneous magnetization), that ultimately determines the magnetic strength of the magnet. This magnetic strength is quantified by the so-called maximum energy product ($BH_{max}$), defined by the area of the largest possible rectangle that fits under the $BH$ curve in the second quadrant, which measures the potential energy stored in the stray field of the magnet.[2]

Figure 1d shows the magnetic hysteresis of an ideal permanent magnet, in which all magnetic spins are perfectly aligned (and therefore, $M_r = M_s$) but in real magnets, the remanence value is smaller than the saturation (*i.e.*, $M_r < M_s$). It follows that, as the $M_r$ value approaches $M_s$, the loop turns more squared, and in turn, $BH_{max}$ is maximized. Hence, the squareness and magnetic alignment is often measured in terms of in $M_r/M_s$ ratio,[3] which is another of the key parameters to be improved for permanent magnets.

## 1.2 Materials for permanent magnets: Current status

Magnetic materials have the unique ability to directly interconvert between electrical and mechanical energy. A moving magnet can induce an electric current to generate electrical energy, and oppositely, an electric current can be used to generate a magnetic field and exert a magnetic force. These electromagnetic properties underpin the operation of electric generators and motors, making magnetic materials critical for the transition towards an environmentally friendly and sustainable future.[2] As a result, the worldwide permanent magnet market is expected to reach $39.71 Billion by 2030, according to the 8.6% compound annual growth rate (CAGR) forecast in the last Grand View Research report.[4]

Figure 2a illustrates the relative performance in terms of $BH_{max}$ and $H_c$ for the most important families of commercial PM materials, including AlNiCo alloys, hard ferrites ceramics, $Nd_2Fe_{14}B$ and $SmCo_5$. The high-performance ($BH_{max} > 100$ kJ/m³) permanent magnet market is currently dominated by the rare earth element (REE)-containing materials $Nd_2Fe_{14}B$ (strongest magnet) and $SmCo_5$ (best high temperature performance) due to their superior energy products,[1,5] which is a critical parameter for the performance in applications where miniaturization is a major driving force (*e.g.*, electric vehicle motors, direct-drive generators, electro-acoustic devices, accessory electric motors, mobile phones, sensors, portable electronics, *etc.*). Unfortunately, the use of REE-based materials entails various problems. The compounds rely on scarce REE such as neodymium, samarium or dysprosium, which are classified as critical raw materials, not only owing to their supply risk and price volatility, but also to the harmful environmental impact of their extraction.[6] China has been the undisputed leader in REE mining and production for the last 40 years,[7] and despite other countries attempting to gain ground, today China still accounts for more than 60% of the world REE production.[8] Consequently, over the last 20 years geopolitical circumstances have often led to erratic price fluctuations. Furthermore, the cobalt used in $SmCo_5$ magnets is another problematic element. The supply chains for the bulk part (>50%) of the cobalt used in advanced materials can be traced back to the cobalt mines in the Democratic Republic of the Congo, where artisanal miners (including thousands of children) work under extremely hazardous conditions.[9] As a consequence, the development of REE-poor or REE-free alternatives has long been an important research topic in the PM field.

Although the undisputed strength of REE-magnets is necessary for the highest-performance applications, there are many other applications that are less demanding in terms of magnetic strength, where a compromise (see Figure 2b) must be made between other factors such as price, stability, processability, *etc*.[10] At this end of the spectrum, hard ferrite magnets have long been the material of choice for lower grade applications (<25 kJ/m³). However, as illustrated by the arrow in Figure 2a, a considerable performance gap exists in the intermediate performance range between the cheaper AlNiCo and hard ferrite PMs and REE PMs. Consequently, for many applications it is often necessary to use an expensive and excessively strong REE magnet, in lack of an intermediate alternative. Here, a modest performance improvement of lower grade magnets would be sufficient to replace REE PMs while remaining within a weight range suitable for the application.

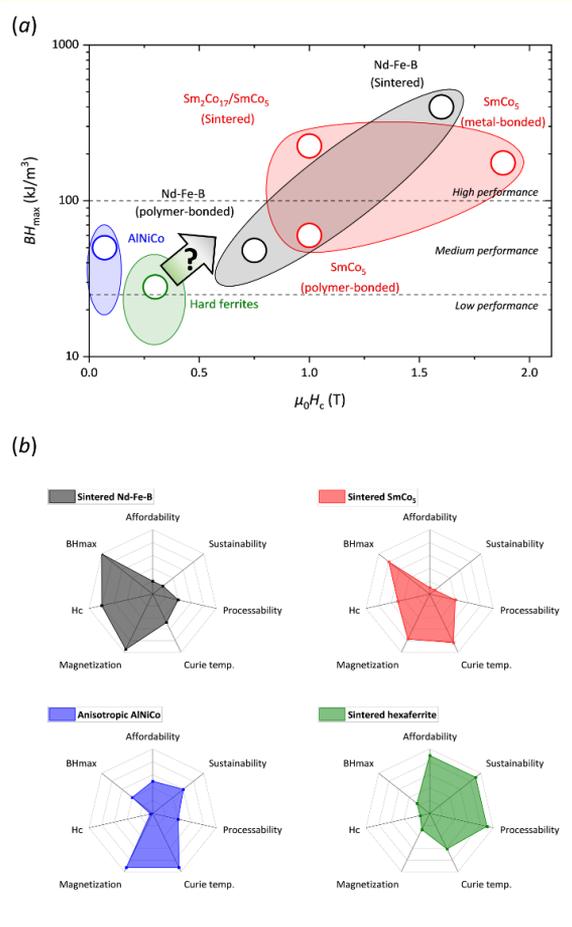

Figure 2: (a) Diagram of $BH_{max}$ vs coercive field for the main families of commercially available hard magnetic materials. (b) Radar plots of key extrinsic properties of sintered $Nd_2Fe_{14}B$, sintered $SmCo_5$, anisotropic AlNiCo and sintered hexaferrite magnets. Figures based on values from [11].

In this context, hexaferrites have long been considered good candidates for replacing REE magnets in the intermediate performance range, due to their reasonably good performance, high Curie temperature (>700 K) and excellent chemical stability, which all comes at a fraction of the cost of REE magnets.[12,13] In fact, hard ferrites are the most produced magnetic material, despite their moderate performance compared to REE magnets.[14] In 2013 they were reported to account for 85 % of the total PM market by manufactured mass, although they only represented 50 % of the market by sales.[15]

While recent studies have demonstrated new approaches to improve magnetic properties of hard hexaferrite powders (*e.g.* nanostructuring,[16−19] chemical substitution, [20−23] exchange spring composites [19,24,25]), manufacturing dense sintered pellets of sufficient structural integrity without degrading the optimized properties has proven a key challenge. In practice, this prevents the replacement of expensive and unsustainable REE PMs in a range of applications, and is the reason why hard ferrites still generate great scientific interest.[26] The present chapter aims at summarizing the most relevant recent achievements and progress in the field, as well as key challenges encountered during the fabrication and sintering of dense ferrite magnets.

## 2. Hard ferrites: M-type hexaferrites

### 2.1 Crystal and magnetic structure

The so-called hexaferrites, hexagonal ferrites or simply hard ferrites, are a family of ternary or quaternary iron oxides with hexagonal crystal lattice of long unit cell *c*-axis (≈23−84 Å).[26] Of the materials in the hexaferrite family, the M-type hexaferrites have been widely used for application as permanent magnets. With chemical formula $M\mathrm{Fe_{12}O_{19}}$ ($M = \mathrm{Sr^{2+}}$ or $\mathrm{Ba^{2+}}$), the Sr and Ba M-type ferrites (SrM and BaM) are isostructural and exhibit very similar magnetic characteristics. The compounds have a large uniaxial magnetocrystalline anisotropy and a magnetic easy axis along the crystallographic *c*-direction. This strong intrinsic anisotropy results in a high $H_c$, making them very resistant towards demagnetization (*i.e.* magnetically hard) and therefore attractive as PM materials.

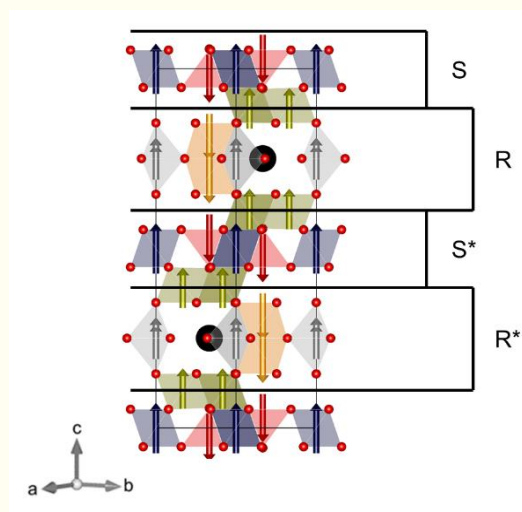

Figure 3. Crystal and magnetic structure of Sr (Ba) hexaferrite. Black and red spheres represent $\mathrm{Sr^{2+}}$ ($\mathrm{Ba^{2+}}$) and $\mathrm{O^{2-}}$ ions. Colored polyhedra illustrate the 5 different crystallographic sites of $\mathrm{Fe^{3+}}$ and arrows symbolize the $\mathrm{Fe^{3+}}$ magnetic spins.

Figure 3 illustrates the crystal and magnetic structures of M-type hexaferrites. They display a hexagonal magnetoplumbite structure (space group $P6_3/mmc$) with very anisotropic unit cell ($a \approx 5.9$ Å, $c \approx 23$ Å). $Fe^{3+}$ ions occupy interstitial positions in a hexagonal close-packed structure of $O^{2-}$ and $Sr^{2+}$ ($Ba^{2+}$) ions.[26–28] With 2 formula units per unit cell (64 atoms), SrM has a crystallographic density of 5.3 g/cm³ (5.1 g/cm³ for BaM).[29,30] The crystal structure may also be described in terms of stacking of simpler structural blocks (cubic S and hexagonal R blocks) which are in turn stacked onto similar blocks rotated 180° about the $c$-axis (S* and R* blocks, respectively).[28]

## 2.2 Magnetic properties

Table 1 compares the intrinsic magnetic properties of SrM and BaM with that of other important magnetic compounds. The theoretical magnetic moments (at 0 K) of the hexaferrite crystal structures can be calculated from the ferrimagnetic ordering of the magnetic $Fe^{3+}$ ions in the structure (see arrows in Figure 3), yielding values of 20.6 $\mu_B$/molecule for SrM and 20 $\mu_B$/molecule for BaM.[26,31] This results in fairly good saturation magnetization, $M_s$, and magnetic induction, $B_s$, values. The Curie temperature, $T_C$, of the M-type hexaferrites is more than 100 °C above that of the much used REE-based $Nd_2Fe_{14}B$ hard phase.

The large uniaxial anisotropy of the hexagonal lattice of SrM and BaM ($c/a = 3.9$) causes a large magnetocrystalline anisotropy along the $c$-axis, which yields relatively high anisotropy constants, $K_1$ (see Table 1)[32–34] and a large theoretical maximum $H_c$ of 594 kA/m.[26] For a hypothetical fully-dense and perfectly-oriented hexaferrite magnet, a theoretical maximum $BH_{max}$ of 45 kJ/m³ has been estimated.[1]

Table 1. Intrinsic magnetic parameters at room temperature (RT) for some representative soft and hard magnetic phases. Data extracted from [34] unless otherwise stated.

| | $M_s$ (Am²/kg) | $B_s$ (T) | $T_C$ (K) | $K_1$ (MJ/m³) |
|---|---|---|---|---|
| $Fe_{0.65}Co_{0.35}$ | 240 | 2.45 | 1210 | 0.018 |
| Fe | 217 | 2.15 | 1044 | 0.048 |
| $AlNiCo5$ [1] | 159 | 1.40 | 1210 | 0.68* |
| $CoFe_2O_4$ [11] | 75 | 0.5 | 793 | 0.27 |
| $BaFe_{12}O_{19}$ | 72 | 0.48 | 740 | 0.33 |
| $SrFe_{12}O_{19}$ | 72 | 0.48 | 746 | 0.35 |
| $Nd_2Fe_{14}B$ | 165 | 1.61 | 588 | 4.9 |
| $SmCo_5$ | 100 | 1.07 | 1020 | 17.2 |
| $Sm_2Co_{17}$ | 118 | 1.25 | 1190 | 4.2 |

*shape anisotropy

## 3. Sintered hard ferrite permanent magnets

Towards the effective implementation of permanent magnets into a device, the material in powder form has to be compacted into dense, mechanically stable and magnetically-oriented pieces (*i.e.*, magnets). This conforming/densification process (called sintering) generally involves applying elevated pressures and/or temperatures to the material in powder shape.[35] As for most other materials, the mechanical properties of the sintered piece relies on a high density. However, the importance of achieving a highly dense magnet is enhanced for PMs, since the magnetic performance ($BH_{max}$) is measured per volume unit, and hence, it is directly proportional to the density.

The high sintering temperatures often end up undesirably altering the functional properties of the starting material and therefore, great efforts are dedicated to both (i) adapting the sintering methods to the specific material of interest and (ii) developing novel sintering strategies that lower the working temperatures, aiming at minimizing the damage.[35] In the particular case of hexaferrites, a common problem is the formation of hematite ($\alpha$-$Fe_2O_3$) as a side phase. This iron oxide is very prone to appear, as a result of its high stability, and causes a decrease of saturation magnetization, due to the antiferromagnetic nature of the phase. Fortunately, it has been shown that $\alpha$-$Fe_2O_3$ can be avoided when the starting $M$Fe$_{12}$O$_{19}$ powders have the right $M$:Fe stoichiometry, yielding $M_s$ values approaching the expected $\approx 70$ Am²/kg.[26] In contrast, limiting the grain growth to circumvent the detrimental impact on $H_c$ has proven more challenging.[36] Thus, M-type ferrites in powder form often present coercivities far below the theoretical value, and the situation worsens for sintered pieces (see Table 7 in ref. [26] for an extensive sample record). Owing to this, sintered hexaferrite magnets are generally inferior to the theoretically achievable 45 kJ/m³,[1] although specific studies have managed to come fairly close to this value.

Another important aspect in the sintering of PMs is the magnetic alignment of the constituent particles and domains in the material. The magnetic particles may (or may not) be magnetically aligned, resulting in anisotropic (isotropic if not aligned) magnets. The greater the magnetic alignment, the more the $M_r$ value approaches $M_s$, yielding a more square-shaped $MH$ curve (as illustrated by the black curve in Figure 1c), thereby maximizing $BH_{max}$. Thus, the $BH_{max}$ of mass-produced isotropic M-ferrite magnets is around 10 kJ/m³, while the anisotropic kind ranges from 33 to 42 kJ/m³.[37–40]* The magnetic alignment has been traditionally carried out by application of an external magnetic field,[26,41] although recently patented methods have succeeded in suppressing the external field by taking advantage of the shape of the particles.[42,43] Notably, the M-type ferrites are prone to form platelet-shaped particles, with magnetization direction parallel to the platelet normal vector (see Figure 4a). As illustrated in the figure, the platelet shape of the particles favors magnetic (and crystallographic) alignment upon compaction.

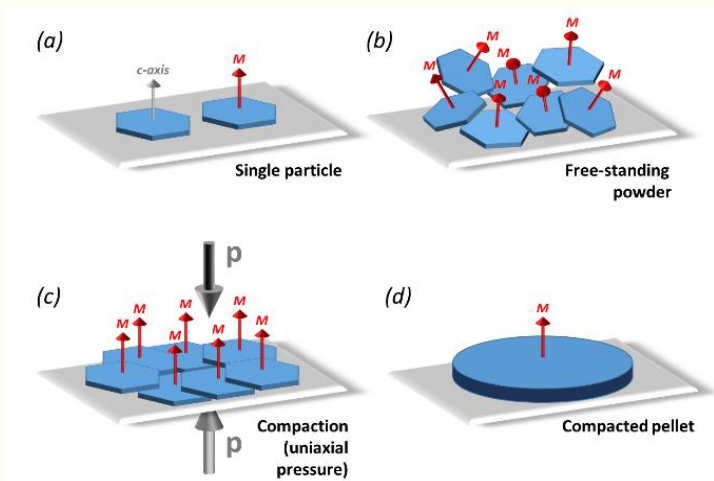

Figure 4. $M$Fe$_{12}$O$_{19}$ particles, displaying typical hexagonal platelet shape with the easy axis of magnetization normal to the platelet plane (and parallel to the crystallographic *c*-axis). This shape favors magnetic (and crystallographic) alignment upon application of uniaxial pressure. Adapted with permission from [13]

---

* Ferrite magnets with higher $BH_{max}$ values are available commercially (up to 44 kJ/m³), but in those cases the material is doped with *e.g.*, La or Co.[39,40]

During the last decades, different sintering strategies have been investigated aiming at maximizing both the magnetic alignment (boosting $B_r$ and $M_r/M_s$) and the $H_c$ on the sintered material. Lately, efforts have also been devoted to making the processes greener and increasing recycling rates. The following sections intend to offer an overview of the pros and cons of each of the alternatives.

## 3.1 Conventional sintering

Hexaferrites were first developed as a PM material by researchers at the Philips Research Laboratories in 1950s. In 1952, Went *et al.* prepared a Ba-ferrite magnet with a good $H_c$ value ($\approx$240 kA/m), although a limited $B_r$ derived from its isotropic nature (0.21 T) yielded a modest $BH_{max}$ of 6.8 kJ/m³.[44,45] Two years later, Stuijts *et al.* developed a conventional sintering (CS) strategy to produce anisotropic BaM magnets with $BH_{max}$ up to 28 kJ/m³,[41] which is essentially the method used nowadays to make sintered ferrite magnets industrially. In brief, a sludge of BaM powders and water is compacted while being held it in an external magnetic field, producing a consolidated piece (still poor in density) which is subsequently sintered at temperatures above 1100 °C to promote densification. Stuijts *et al.* explored sintering temperatures between 1250 and 1340 °C and noted that increasing the temperature maximizes the density and the magnetic alignment (and therefore $B_r$), but at the cost of decreasing $H_c$, as a consequence of the grain growth promoted by the elevated temperatures. This problem, encountered already in 1954, has been subject of extensive research since.

As mentioned earlier, structural characteristics such as crystallite size, size distribution and crystallite morphology can largely affect the coercivity of ferrite magnets. In particular, highest $H_c$ values are attained for crystallite sizes close to the critical single-domain size defined earlier.[33,45,46] The difficulty not only lies in being able to produce particles of a specific size in a controlled manner, but it begins with determining what this critical size is for a specific material. For isotropic SrM crystallites, the critical single-domain size has been estimated to be around 620–740 nm.[16,47] However, the experimentally reported crystallite/particle single-domain sizes of SrM span from 30 nm all the way up to 830 nm.[47] This is due to the high influence of particle morphology in the attained coercivity, as well as to the different characterization methods used to determine the reported size (*i.e.* particle *vs.* crystallites, number *vs.* volume weighted, *etc.*). A study by Gjørup *et al.* showed that a much smaller critical single-domain size is obtained for highly anisotropic crystallites, and therefore not only the overall size, but also the aspect ratio of anisotropic SrM crystallites should be considered when trying to maximize $H_c$.[47]

Notably, reducing the size of the starting powders does not necessarily yield to better coercivities, as the grain growth upon sintering seems to be even greater when dealing with materials of smaller particle sizes.[48–50] El Shater *et al.* sintered nanometric BaM (100–200 nm) at 1000 and 1300 °C, producing average particle sizes of 0.537 and 16.35 μm, respectively, with the consequent drop in coercivity (from 271 to 56 kA/m) and the $M_r$ gain.[51] Therefore, the choice of sintering temperature must be a compromise between minimizing grain growth (to maximize $H_c$) and maximizing densification (and in turn, $M_r$).

A common approach for limiting grain growth has been the use of sintering additives. Kools proposed a mechanism through with $SiO_2$ would prevent the growth of SrM grains during sintering and proved the effect for a range of $SiO_2$ concentrations (0.36–1.44 wt.%).[52,53] Beseničar *et al.* reported that, besides limiting the growth, $SiO_2$ induces some ordering of the SrM particles, resulting in very anisotropic magnets with high relative density (97%) and satisfactory magnetic properties ($B_r \approx$ 0.39 T, $H_c \approx$ 340 kA/m).[54] Kobayashi *et al.* determined the optimal $SiO_2$ concentration to be between 1 and 1.8 wt.%, showing a detrimental effect on $H_c$ for greater $SiO_2$ additions.[55] Guzmán-Mínguez *et al.* reported the appearance of ≈20 wt.% α-$Fe_2O_3$ as a secondary phase for $SiO_2$ concentrations>1 wt.%.[56]

CaO has been reported to favor densification, and therefore, it has also been explored as a sintering additive for hexaferrites, in this case with the aim of boosting $M_r$, although at the expense of aggravating the grain growth effect.[46,55,57] In this context, the combined use of both additives has also been investigated. Lee *et al.* reported a decent $BH_{max}$ of 29.4 kJ/m³ when adding 0.6 wt.% $SiO_2$ and 0.7 wt.% CaO, but neither remanence nor coercivity were terrific ($B_r$ = 0.36 T, $H_c$ = 281 kA/m).[58] Töpfer *et al.* fabricated a very dense SrM magnet (98%) with a notable $B_r$ value of 0.42 T by incorporating 0.25 wt.% of $SiO_2$ and 0.25 wt.% CaO, although a moderate coercivity value of 282 kA/m only allowed for a $BH_{max}$ = 32.6 kJ/m³.[59] Huang *et al.* tested the combined addition of $CaCO_3$, $SiO_2$ and $Co_3O_4$ (1.1, 0.4 and 0.3 wt.%, respectively), managing a remarkable $BH_{max}$ of 38.7 kJ/m³, owing to an exceptional remanence (0.44 T) and despite a modest coercivity (264 kA/m).[60]

Slightly superior magnetic parameters ($B_r$ = 0.44 T, $H_c$ = 328 kA/m, $BH_{max}$ = 37.6 kJ/m³) have been obtained by from a two-step sintering (TSS) method adapted to SrM by Du *et al.*.[61] Here, the powders were cold-pressed as usual CS, but the subsequent thermal cycle used for sintering was slightly more elaborate: after a first high temperature step, in which the maximum temperature (1200 °C) is maintained for only 10 min, a longer (2 h) heating step at 1000 °C provides for full densification of the SrM magnet.[61] The scanning electron micrograph (SEM) in Fig. 6(e) from ref. [61] illustrates the confined grain size, the high density and high degree of alignment justifying the good magnetic performance. A more recent work by Guzmán-Mínguez *et al.*[62] combined a TSS approach with the addition of 0.2% PVA and 0.6% $SiO_2$, realizing great control of the grain growth at 1250 °C (see Figure 5) although the obtained magnetic properties were not as good as the ones previously reported by Du *et al.*.

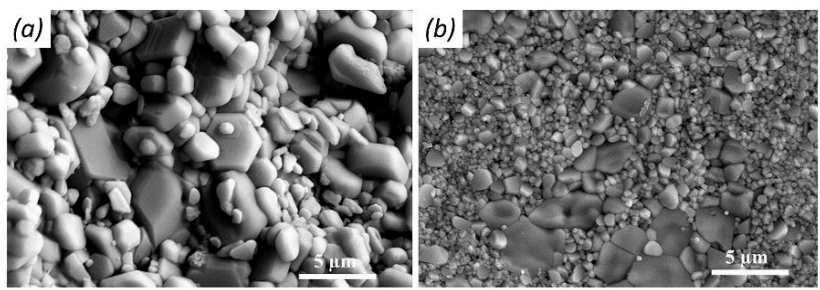

Figure 5. SEM images of SrM pellet sintered at 1250 °C by (a) conventional sintering and (b) two-step sintering. Reprinted from [62], Copyright 2021, with permission from Elsevier.

## 3.2 Spark plasma sintering

In the 1990s, a new commercial apparatus based on resistive sintering, called spark plasma sintering (SPS) was developed by Sumitomo Heavy Industries Ltd. (Japan).[63] The SPS method is based on the use of an electrical current and a uniaxial mechanical pressure under low atmospheric pressure, to simultaneously heat and compact a powder sample.[64] The starting powders are typically loaded in a graphite die, which is placed between two electrodes in a water-cooled vacuum chamber. A uniaxial pressure is applied to the die while passing a DC electrical current through, which heats up the sample due to the Joule effect (see Fig. 1 in ref. [65] for a typical SPS setup). The inventors of the system claimed the generation of plasma to take place, thus leading to the technique's name. However, although it is generally accepted that plasma may be generated between particles due to electrical discharges, there is no conclusive experimental evidence of such occurrence.[64] Therefore, SPS is sometimes referred to by alternative names, such as field-assisted sintering technique (FAST). The simultaneous application of temperature and pressure can also be obtained by conventional hot pressing (HP). However, in SPS and HP, heat is produced and transmitted to the material in different ways. In conventional heating the powders are sintered by heating the entire container using external heating elements in a furnace. This leads to slow heating rates, long sintering times and waste of energy in heating up all the components. The SPS method, however, has allowed increasing the heating rates, lowering the working temperatures and reducing the dwell times.[66,67] These benefits make SPS a good alternative when the goal is to limit the grain growth during sintering,[67] and potentially improve the obtained $H_c$ (and $BH_{max}$) values of sintered hexaferrite magnets.

Numerous investigations focusing on sintering hexagonal ferrites by SPS have been published in the last two decades. Obara *et al.* prepared fully-dense SrM magnets by SPS at 1100 °C and 50 MPa for only 5 min.[65] A fairly competitive $H_c$ of 325 kA/m was obtained by doping with $La_2O_3$ (1 wt.%) and $Co_3O_4$ (0.1 wt.%). Although the measured hysteresis loops were rather squared, the remanence value (0.32 T) was not sufficient to guarantee a noteworthy energy product ($BH_{max}$ = 18.3 kJ/m³). Mazaleyrat *et al.* sintered BaM nanopowders with sizes below 100 nm and managed to hold grain growth and produce a $H_c$ of 390 kA/m,[68] which even surpasses the value reported for the La and Co-doped material described above. Unfortunately, a deficient density (88%) degraded the $BH_{max}$ down to 8.8 kJ/m³. Ovtar *et al.* sintered the same batch of 90 nm BaM nanoparticles by both CS and SPS, producing much smaller sizes through the second method.[69] Additionally, they realized that secondary phases ($Fe_3O_4$, $\alpha$-$Fe_2O_3$) tend to form on the surface of the BaM SPS pellets, and tested different materials for the protective discs separating the sample from the graphite die (BN, Au, $\alpha$-$Al_2O_3$) concluding that $\alpha$-$Al_2O_3$ was the one performing best. The resulting density was rather low 82% but the coercivity was adequate (350 kA/m). Stingatiu *et al.* attempted downsizing a μm-sized SrM material by a ball-milling step prior to consolidation through SPS.[70] The resulting density was satisfactory (90%) but unfortunately, ball-milling was seen to amorphize the surface of the SrM, which triggered formation of secondary phases during SPS, this having a detrimental effect on the magnetic properties ($BH_{max}$ < 10 kJ/m³).

Saura-Múzquiz *et al.* prepared nm-sized SrM powders by hydrothermal synthesis (HT) with hexagonal plate-like particles (such as those in Figure 4) with very small sizes; in some cases, the platelets were as thin as a single unit cell (*i.e.* <3 nm).[17] These HT-synthesized SrM powders were consolidated by SPS yielding appropriate $H_c$ values of 301 kA/m. More importantly, the highly anisotropic shape of the particles provided for a pronounced magnetic alignment of the sintered SrM magnets, inherently occurring as a result of simultaneous application of elevated temperature and uniaxial pressure, just as illustrated in Figure 4. Here, an $M_r/M_s$ ratio of 0.89 was reached without applying an external magnetic field neither before nor during sintering, yielding a $BH_{max}$ value of 26 kJ/m³. Figure 6a shows the magnetic hysteresis of the HT powders and the corresponding SPS pellet, evidencing the squareness of the latter. Achieving magnetic alignment without a magnetic field is very convenient from an industrial point of view, because it allows a full step to be removed from the manufacturing process (*i.e.*, the magnetization), which simplifies the procedure, reduces costs and increases energy efficiency.[42] Figure 6b displays the powder X-ray diffraction (PXRD) data measured on both SrM powders and SPS pellet. Despite the very dissimilar appearance, Rietveld analysis demonstrates that both PXRD patterns are consistent with pure-phase $SrFe_{12}O_{19}$ although with notable differences in crystallite size and orientation. The highly anisotropic shape of the powders is visible from the sharpness of the *hkl*-reflections describing the crystalline on the platelet plane, such as (110) or (220), compared to the large broadening of those associated to the platelet thickness, *e.g.*, (008), all this in agreement with much smaller sizes along the *c*-axis than on the *ab*-plane (*i.e.*, thin platelets). Regardless of the difference in peak broadening, Bragg reflections of all orientations are present in the PXRD pattern measured for the SrM powders, demonstrating a random orientation of the crystallites. However, the very intense *hh0* reflections are absent for the PXRD pattern recorded for the SPS pellet, while 00*l* reflections (as well as others with high contribution from the *c*-crystallographic direction) are systematically intensified, thus indicating a marked preferred orientation of the platelets. As explained before, for M-type platelet-shaped particles, crystallite/particle alignment goes together with magnetic alignment. The crystallographic alignment was further studied based on pole figure measurements (Figure 6c), a slightly more complex diffraction measurement enabling quantification of the degree of orientation (Figure 6d).

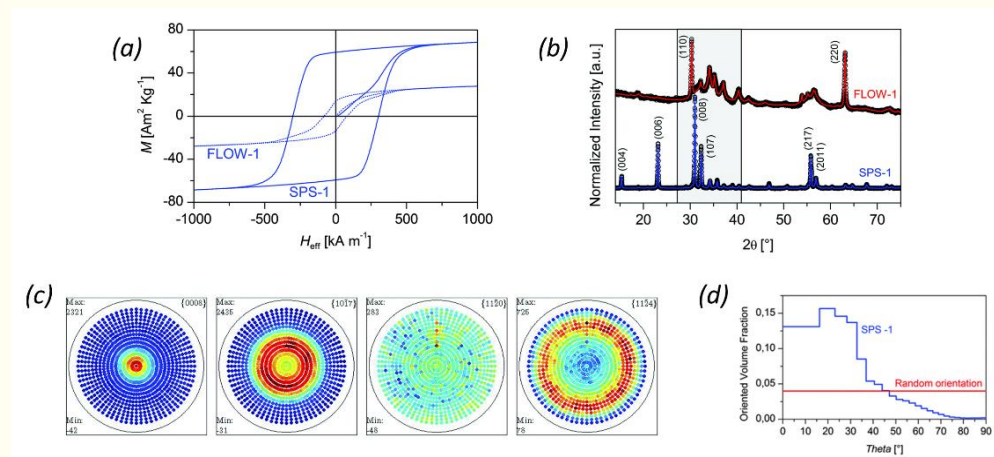

Figure 6. (a) Magnetic hysteresis loop of HT-synthesized SrM nanoparticles and corresponding SPS pellet. (b) PXRD data along with Rietveld model of the same samples. (c) X-ray pole figure measurements and (d) oriented volume fraction of SPS pellet. Reproduced from Ref. [17] with permission from the Royal Society of Chemistry.

Optimization of both the HT synthesis route[71] and the SPS protocol[18,72] as well as correlation of structural and magnetic properties, allowed reaching $M_r/M_s$ ratios as high as 0.95, although at the cost of reducing $H_c$ down to 133 kA/m, with which the $BH_{max}$ improvement was only moderate (29 kJ/m³).[73] However, performing a thermal treatment at 850 °C after SPS was enough to reach a $BH_{max}$ = 36 kJ/m³, value on the order of the highest-grade commercially available ferrite magnets,[37–40] while avoiding the use of an external magnetic field. Applying this SPS protocol to SrM powders produced by synthesis methods other than HT did not yield such outstanding magnetic properties, due to an inferior particle orientation degree and, hence, a poorer magnetic alignment.[72,74] A newer study by Saura-Múzquiz *et al.* confirmed that the degree of magnetic alignment using this preparation method could be tuned by modifying the aspect ratio of the initial powders, reaching almost fully-aligned pellets ($M_r/M_s$ = 0.9) with densities above 90%.[74] Higher alignment leads to higher squareness and thus greater $M_r/M_s$ ratio and $BH_{max}$, but it is accompanied by a reduction in $H_c$ due to the inversely proportional relationship that exists between magnetization and coercive field. Nonetheless, by reducing the degree of alignment they were able to obtain SrM magnets with a large $H_c$ of 412 kA/m, proving the potential of SPS to overcome the reduction of $H_c$ due to excessive crystallite growth.

Recently, Vijayan *et al.* reported the use of SPS not only for densification of ferrite powders, but for the direct synthesis of aligned SrM magnets.[75–77] In this study, SrM is synthesized directly during the SPS process, using a precursor powder of antiferromagnetic six-line ferrihydrite (FeOOH) platelets mixed with SrCO$_3$. A low SPS temperature of ≈750 °C was sufficient to drive the reaction between the six-line phase and SrCO$_3$ to produce SrFe$_{12}$O$_{19}$, while the anisotropic shape of the hydrothermally synthesized six-line phase ensured the alignment of the resulting SrM particles. Following this synthesis method, they were able to produce a dense SrM magnet with a $BH_{max}$ of 33(4) kJ/m$^3$, a $M_r/M_s$ of 0.93 and a $H_c$ of 247 kA/m.

## 3.3 Microwave sintering

In the field of hexaferrite research, microwaves (MWs) have mainly been used for synthesis purposes although a few sintering attempts using MWs have also been reported.[78–80] In all of them, powders are initially cold-pressed followed by a MW treatment, using frequencies in the GHz range, to sinter the piece.[35] In 1999, Binner *et al.* used MWs to sinter ferrite nanoparticles reporting a limited grain growth for non-agglomerated starting powders, although they failed to avoid cracks in the final sintered pieces.[78] Ten years later, Yang *et al.* succeeded in making 97 % dense BaM magnets by MWs sintering.[79] They also managed to prevent the appearance of α-Fe$_2$O$_3$ in the final material, although they did not succeed in preventing grain growth, in turn producing a rather low $H_c$ (<50 kA/m).

Recently, Kanagesan *et al.* tested fast heating rates (50 °C/min) and short dwell times (10 min) to MW sinter some Sr-ferrite powders synthesized by sol-gel.[80] The MW sintering at 1150 °C yielded a 95 % SrM ceramic magnet with a fairly high $H_c$ of 445 kA/m. However, the $M_s$ value (50 Am$^2$/kg) was not outstanding, although the sample seems relatively pure from powder diffraction data. The $M_r$ value is also rather low (≈30 Am$^2$/kg), which is expected from the poor alignment of the SrFe$_{12}$O$_{19}$ particles observed in the corresponding SEM micrograph (see Fig. 2 in ref. [80]).

## 3.4 Cold sintering process

In 2016, Guo *et al.* reported a new sintering strategy named cold sintering process (CSP), with which they were able to attain high densification degrees for a wide range of inorganic materials at temperatures ≤ 200 °C, fabricating materials with functional properties comparable to those made by conventional high-temperature approaches.[81] For CSP, the ceramic powders are mixed with a small amount of aqueous solution which partially dissolves the particle edges and facilitates diffusion and mass transport, aiding the sintering process, which in turn occurs at lower temperatures. Sintering at low temperatures is very attractive in general, as it reduces the energy demands, making the process greener and more cost-efficient. This is especially interesting for M-ferrite magnets, as lower working temperatures are expected to minimize grain growth. The exact role of the solvent during CSP is still under discussion, but it is believed to induce the formation of an amorphous phase at the grain boundaries which eases sintering and may also restrict grain growth.[81]

To the best of our knowledge, there is only one research group which has tested CSP on hard ferrites. In 2021, Serrano *et al.* patented a CSP method that allows fabrication of dense SrM magnets with magnetic properties in the order of medium-grade commercial ferrite magnets.[43] In the CSP method developed by Serrano, SrM powders are mixed with glacial acetic acid and the wet mixture is subjected to a uniaxial pressure (≈400 MPa) while heated at 190 °C.[82] After CSP, relative densities of about 85% are obtained, which can be driven up to 92% by subsequently treating the sintered piece at 1100 °C for 2 h. This last sintering step also has a beneficial effect on the magnetic properties (see Figure 7A). In particular, $M_s$ at 5 T increases from 49.2 to 73.7 Am²/kg and $H_c$ goes from 119 to 223 kA/m. For the final product, a $M_r/M_s$ ratio of 0.68 was obtained. The density obtained by conventional sintering at 1100 °C for 4 h (no solvent, no hot compression) was only 77% and the magnetic properties slightly inferior (see Figure 7B). Conventional sintering at 1300 °C yielded higher density (97%) but very poor magnetic properties ($H_c$ = 48 kA/m, $M_r/M_s$ = 0.33), due to the dramatic grain growth caused by the high temperature (see bottom FE-SEM micrograph from Figure 7B).

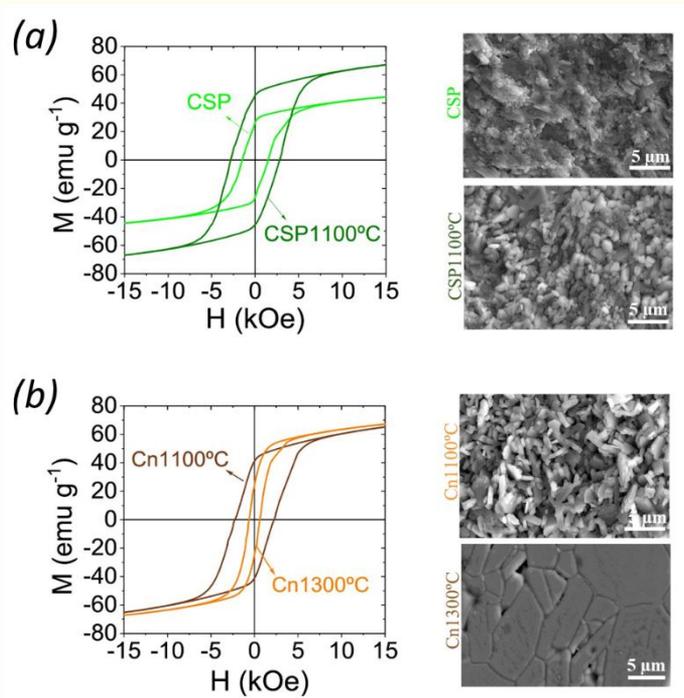

Figure 7. Magnetic hysteresis and FE-SEM corresponding to SrFe$_{12}$O$_{19}$ magnets fabricated by A) CSP at 190 °C, CSP followed by annealing at 1100 °C, B) conventional sintering at 1100 and 1300 °C. Reprinted from [82] with permission from Elsevier.

Further investigations have been carried out using different organic solvents (*i.e.*, oleic acid, oleylamine) and widening the pressure and temperature ranges explored (up to 270 °C and 670 MPa).[83] In all cases, the average grain size of the CSP ceramic magnet was about 1 μm (similar to the starting SrM powders) while similar conventional sintering processes typically yield average grain sizes above 3 μm.[62]

With the aim of further improving the density and magnetic properties of CSP magnets, the addition of a small amount (10 wt.%) of nanometric SrM to the original micrometric SrM powders was tested, moderately increasing $H_c$ (239 kA/m) and $M_r/M_s$ (0.73), although the density value continued at 92%.[84] These numbers are competitive in the context of commercial SrM magnets. As an example, the Hitachi's NMF-7C series display values of $H_c$ = 220–260 kA/m and $M_s$ = 68 Am²/kg).[40]

## 4. Summary and perspective

In the present chapter, the main sintering approaches applied to manufacturing hard ferrite ceramic magnets have been reviewed. Table 2 summarizes the properties of top $SrFe_{12}O_{19}$ magnets fabricated by the various discussed sintering strategies. Conventional sintering (CS) continues to be the quintessential industrial method for M-type hexaferrite PM fabrication, owing to its technical simplicity and the relatively good resulting properties. However, this approach is highly inefficient, as most of the energy employed is irreversibly dissipated as heat.[85] Therefore, the search for more energy-efficient methods continues to be an active field of research.

Table 2. Magnetic parameters and relative density for top representatives of $SrFe_{12}O_{19}$ magnets manufactured following the different sintering approaches described in the present chapter, *i.e.*, conventional sintering (CS),[61] spark plasma sintering (SPS),[18] cold sintering process (CSP),[84] and microwave sintering (MWs).[80]

| | $M_s$ (Am²/kg) | $M_r/M_s$ | $M_r$ (Am²/kg) | $H_c$ (kA/m) | $BH_{max}$ (kJ/m³) | $\rho_{rel}$ (%) |
|---|---|---|---|---|---|---|
| **CS** | ≈68 | ≈1 | 68 | 328 | 37 | ≥99% |
| **SPS** | 73 | 0.93 | | 225 | 36 | >95% |
| **CSP** | 73 | 0.73 | | 239 | – | 92% |
| **MWs** | 50 | ≈0.62 | | 445 | – | 95% |

*Approximate values (≈) are graphically estimated from the article figures.

Multiple studies have demonstrated that spark plasma sintering (SPS) allows production of PMs with much higher $M_r/M_s$ ratios than CS. However, the increase in texture comes at a cost of reduction in $H_c$ values, which therefore still need to be improved. As a result, magnets made using SPS end up displaying a similar performance ($BH_{max}$) to the best CS examples. Additionally, technical challenges hinder the replacement of CS by SPS in the industrial production of magnetic ferrites, since current SPS machines only allow producing relatively small pieces with very few specific shapes (typically cylindrical pellets).

Only a few attempts have so far been made to densify SrM by the relatively new cold sintering process (CSP) and therefore, there is still much to explore and optimize. However, the CSP has already allowed preparation of hexaferrite magnets with magnetic properties comparable to medium-high grade commercial ferrites, while lowering the sintering temperature. This reduces the energy consumption by about 9 kWh/kg, which leads to energy savings of ≈29% compared to the sintering methods employed industrially at present.

The results obtained by microwave sintering (MWs) have been very satisfactory in terms of both density and $H_c$, but the resulting $M_s$ and $M_r/M_s$ values are still insufficient to be commercially competitive. As with CSP, reports are scarce and further exploration is required.

Sintering has undergone significant innovation over the last decade,[35] with the introduction of a number of new sintering technologies, such as flash sintering,[86,87] and various modified SPS methodologies, like flash SPS (FSPS),[88] deformable punch SPS (DP-SPS),[89] or cool-SPS.[90] As a result, there are more alternatives available for sintering ferrites with enhanced magnetic characteristics and microstructure. To our knowledge, none of the just mentioned have yet been examined on hard hexagonal ferrites, leaving lots of room for additional study in this area.

## Acknowledgments

C.G.-M. acknowledges financial support from grant RYC2021–031181-I funded by MCIN/AEI/10.13039/501100011033 and by the "European Union NextGenerationEU/PRTR". M.S.-M. acknowledges the financial support from the Comunidad de Madrid, Spain, through an "Atracción de Talento Investigador" fellowship (2020-T2/IND-20581). H.L.A acknowledges the financial support from The Spanish Ministry of Universities (Ministerio de Universidades) and the European Union—NextGenerationEU through a Maria Zambrano—attraction of international talent fellowship grant.